\documentclass[12pt]{article}
\usepackage{amsmath,amssymb,amsfonts,epsf}
\usepackage[nosort]{cite}
\usepackage[margin=1in]{geometry}

\newcommand{\fft}[2]{{\frac{#1}{#2}}}

\def\nn{\nonumber}
\let\bm=\bibitem
\newcommand{\be}{\begin{equation}}
\newcommand{\ee}{\end{equation}}

\def\fft#1#2{\frac{#1}{#2}}
\def\td{\tilde}

\newcommand{\bea}{\begin{eqnarray}}
\newcommand{\eea}{\end{eqnarray}}

\begin{document}

\begin{center}\ \\ \vspace{60pt}
{\Large {\bf The Poynting-Robertson Effect on Solar Sails}}\\ 
\vspace{30pt}
Roman Ya. Kezerashvili and Justin F. V\'azquez-Poritz
\vspace{20pt}

{\it Physics Department\\New York City College of Technology, The City University of New York\\ 300 Jay Street, Brooklyn NY 11201, USA}\\
\vspace{10pt}
{\it The Graduate School and University Center, The City University of New York\\ 365 Fifth Avenue, New York NY 10016, USA}

\vspace{20pt}
{\tt rkezerashvili@citytech.cuny.edu}\\ 
{\tt jvazquez-poritz@citytech.cuny.edu}

\end{center}

\vspace{30pt}

\centerline{\bf Abstract}

\noindent We consider a special relativistic effect, known as the Poynting-Robertson effect, on various types of trajectories of solar sails. Since this effect occurs at order $v^{\phi}/c$, where $v^{\phi}$ is the tangential speed relative to the sun, it can dominate over other special relativistic effects, which occur at order $v^2/c^2$. While solar radiation can be used to propel the solar sail, the absorbed portion of it also gives rise to a drag force in the tangential direction. For escape trajectories, this diminishes the cruising velocity, which can have a cumulative effect on the heliocentric distance. For a solar sail directly facing the sun in a bound orbit, the Poynting-Robertson effect decreases its orbital speed, thereby causing it to slowly spiral towards the sun. We also consider this effect for non-Keplerian orbits in which the solar sail is tilted in the azimuthal direction. While in principle the drag force could be counter-balanced by an extremely small tilt of the solar sail in the polar direction, periodic adjustments are more feasible.


\thispagestyle{empty}

\newpage

\section{Introduction}

Solar electromagnetic radiation can be transmitted, reflected, absorbed and emitted by re-radiation by a solar sail. It is well known that the reflected, absorbed and emitted portions of the radiation can be used to propel the solar sail, due to the force of the electromagnetic pressure. What is less known is that the absorbed and emitted portions of the radiation induce a drag force on the solar sail, thereby diminishing its tangential speed relative to the sun. While this force is relatively small, it can have a long-term cumulative effect on the trajectories of solar sails. In particular, the effect of this drag force can be rather dramatic on the heliocentric distance of solar sails in long-range escape trajectories, while solar sails in bound orbits will spiral inwards to the sun.

The aforementioned drag force on solar sails is associated with the Poynting-Robertson effect, which was first investigated for small spherical particles by Poynting in 1904 \cite{Poynting}. Although it is now realized that this is a special relativistic effect associated with the finite speed of light, Poynting's paper was actually published a year before Einstein's paper on special relativity. This effect was reconsidered in 1937 by Robertson as a manifestly special relativistic effect \cite{Robertson}. Then starting from a paper by Wyatt and Whipple \cite{Wyatt and Whipple}, the Poynting-Robertson effect was used for modeling the orbital evolution of dust particles orbiting the sun and, in particular, accounted for the drag force which causes dust particles to slowly spiral inward towards the sun. Various refinements of the original analysis have been made, such as including the leading-order effect of curved spacetime around the sun for particles of various shapes\footnote{See \cite{Kocifaj} and \cite{Donato} and references therein.}.

While there has been much research devoted to the influence of the Poynting-Robertson effect on the motion of dust grains, the Poynting-Robertson effect was only recently considered for solar sails in bound orbits \cite{Roman Justin}. In this paper, we consider the Poynting-Robertson effect for solar sails in escape trajectories and further elaborate on the cases of bound heliocentric and non-Keplerian orbits. From the rest frame of the solar sail, this effect is associated with the nonradial component of the absorbed solar radiation, due to the relative motion between the sun and the solar sail. From the rest frame of the sun, this results from the solar sail absorbing solar radiation and then emitting it in the forward direction relative to its motion. We would like to mention that there is a second effect that is also due to the absorption of solar radiation, known as the Yarkovsky effect \cite{Opik}, which is related to the emissivity of the solar sail material. For a solar sail that has tangential motion, both types of effects can occur simultaneously.

This paper is organized as follows. In section 2, we present the orbital equations for a solar sail orbiting in the plane of the sun, which takes into account the drag from the Poynting-Robertson effect. In section 3, we investigate the diminished heliocentric distance which results from this drag effect on solar sails in escape trajectories. In section 4, we consider how the Poynting-Robertson effect causes an initially circular orbit to slowly spiral in towards the sun. We consider both bound orbits that lie within the plane of the center of mass of the sun, as well as non-Keplerian orbits which lie outside of this plane. Conclusions follow in section 5.

\section{Orbital equations} 

\begin{figure}[ht]
   \epsfxsize=5.5in \centerline{\epsffile{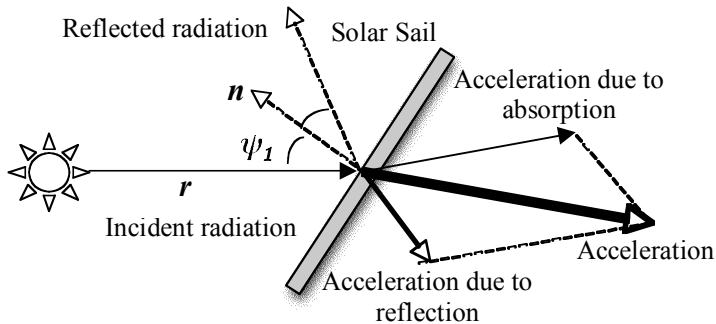}}
   \caption[FIG. \arabic{figure}.]{\footnotesize{Specular reflection leads to an acceleration directed in the opposite direction from the reflected radiation, while absorption leads to an acceleration directed in the same direction as the incident radiation.}}
\label{fig1}
\end{figure}

The total force exerted on the solar sail due to solar radiation is the result
of reflection, absorption and emission by re-radiation. When
the solar electromagnetic radiation interacts with the solar sail material, it
undergoes diffuse as well as specular reflection. The acceleration due to the diffuse reflection is directed
along the normal to the antisun surface area, while the acceleration due to
the specular reflection is directed opposite to the
reflected radiation. In addition, the acceleration produced by the absorption of the solar
radiation is directed along the incident radiation. This is shown in Figure \ref{fig1}.
Moreover, a portion of
the absorbed radiation will be re-radiated from the front and back sides of the
solar sail, which leads to an acceleration normal
to whichever surface has the smaller emissivity. 

Following \cite{mcinnes2,montenbruck}, the
acceleration due to the forces produced by the solar radiation pressure can
be expressed as
\be\label{RK1}
\mathbf{a}=\frac{\kappa }{r^{2}}\cos \psi_1 \left[ a_{1}\widehat{\mathbf{r}}%
-(2a_{3}\cos \psi_1 +a_{2})\widehat{\mathbf{n}}\right] ,\qquad
\kappa \equiv \frac{L_{s}}{2\pi c\sigma },
\ee
where $\widehat{\mathbf{r}}$ is the unit vector in the direction of the
heliocentric vector $\mathbf{r}$, $\widehat{\mathbf{n}}$ is the unit
normal vector for  the surface area facing the sun,  $\psi_1$ is the pitch
angle of the solar sail relative to the heliocentric vector $\mathbf{r}$, $\sigma$ is the mass per surface area of the solar sail, and $L_{S}$= $%
3.842\times 10^{26}$ W is the solar luminosity at 1 AU. The optical
characteristics of the solar sail film $a_{1}$, $a_{2}$, $a_{3}$ are defined
as
\bea \label{RK4}
a_{1} &=&1-\rho s,  \nn\\
a_{2} &=&B_{f}(1-s)\rho +(1-\rho )\frac{\epsilon _{f}B_{f}-\epsilon _{b}B_{b}%
}{\epsilon _{f}+\epsilon _{b}},\\
a_{3} &=&\rho s,  \nn
\eea
where $\rho $ is the reflection coefficient, $s$ is the
specular reflection factor, $\epsilon _{f}$ and $\epsilon _{b}$ are the
emission coefficients of the front and back sides of the solar sail, and $B_{f}$
and $B_{b}$ are the non-Lambertian coefficients of the front and back sides, which describe the angular distribution of the emitted and the diffusely reflected photons. For the case in which the surface of a non-perfectly reflecting solar sail is directly facing the sun, its acceleration is given by
\be
\mathbf{a}=\frac{\kappa }{r^{2}}\left[ a_{1}\widehat{\mathbf{r}}%
-(2a_{3}+a_{2})\widehat{\mathbf{n}}\right] ,  \label{RK5}
\ee

Equations (\ref{RK1}) and (\ref{RK5}) are well known. Now we will bring into consideration an additional force due to absorption which has not been taken into account in these equations. This is because the above description is in the Newtonian framework, for which the solar radiation is purely radial. Thus, for
the case in which the surface of the solar sail is directly facing the sun,
the force due to the solar radiation pressure (SRP) is directed radially
outwards. However, as a special relativistic effect, the solar radiation has
a nonzero tangential component in the frame of reference of the solar sail.
This is due to the relative tangential speed $v^{\phi }=r\dot{\phi}$ between
the solar sail and the sun, where $r$ is the heliocentric distance and $\phi 
$ is the angular coordinate. In particular, in the rest frame of the solar
sail, the solar radiation propagates at an angle $\alpha =\sin ^{-1}(v^{\phi
}/c)$ with respect to the radial direction. Therefore, the absorbed portion
of this radiation leads to a force with a component opposite the direction
of motion. This drag effect is generally known as the Poynting-Robertson effect, which occurs at order $v^{\phi }/c$, and dominates over
other special relativistic effects which are at order $v^{2}/c^{2}$. Note that we are neglecting the redshift in wavelength of the solar radiation due to the radial velocity of the solar sail, which is a higher-order effect.

\begin{figure}[ht]
   \epsfxsize=2.5in \centerline{\epsffile{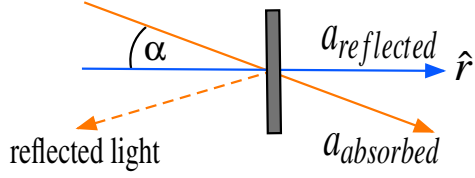}}
   \caption[FIG. \arabic{figure}.]{\footnotesize{Reflected light yields an acceleration in the radial direction, whereas absorbed light leads to an acceleration at an angle $\protect\alpha$ with respect to the radial direction, due to the Poynting-Robertson effect.}}
\label{fig2}
\end{figure}

We will first consider a solar sail whose surface is directly facing the sun and whose motion is restricted to lie within the heliocentric plane. The portion of light reflected by the solar sail leads to a radially outwards acceleration, whereas the portion of light absorbed leads to an acceleration directed at an angle $\alpha$ with respect to the radial direction, as depicted in Figure \ref{fig2}. 

We will be using the reflectivity parameter $0.5\le \eta\le 1$ where $%
\eta=0.5$ ($1$) corresponds to the total absorption (total reflection) of solar radiation by the surface of the solar sail. The fraction of light reflected is 
$2\eta-1$ and the fraction of light absorbed is $2(1-\eta)$. Throughout this paper, we will take a
conservative value of $\eta=0.85$. The acceleration due to the portion of light that is reflected by the solar
sail is directed in the radial direction and has a magnitude of 
\begin{equation}  \label{areflected}
a_{reflected}=\frac{(2\eta-1)\kappa}{r^2}\ \cos^2\alpha.
\end{equation}
We are not considering fluctuations in the solar
radiation and are taking a time-averaged luminosity. One factor of $%
\cos\alpha$ is due to the fact that the \textit{area} of the solar sail is
effectively reduced since it lies at an angle of $\alpha$ with respect to
the direction of the solar radiation. The second factor of $\cos \alpha$
comes from projecting the solar radiation \textit{force vector} along the
direction normal to the surface of the solar sail.

There is also acceleration due to the portion of light that is absorbed,
which is directed at an angle $\alpha$ in the backward direction and has a
magnitude of 
\begin{equation}\label{aabsorbed}
a_{absorbed}=\frac{(1-\eta)\kappa\cos\alpha}{r^2}.
\end{equation}
Projecting the two contributions to the acceleration given by (\ref{areflected}) and (\ref{aabsorbed}) along the radial and angular directions and including the acceleration due to the gravitational field of the sun gives 
\begin{eqnarray}  \label{a2}
a^r &=& \frac{\eta\kappa \cos^2\alpha}{r^2}-\frac{GM}{r^2},  \notag \\
a^{\phi} &=& -\frac{(1-\eta)\kappa\sin\alpha \cos\alpha}{r^2}.
\end{eqnarray}
where the mass of the sun is about $M=1.99\times 10^{30}$ kg.

In polar coordinates, the components of acceleration can be written as 
\bea\label{a1}
a^r &=& \ddot r-r\dot\phi^2,\nn\\
a^{\phi} &=& r^{-1} \partial_t (r^2\dot\phi).
\eea
Equating the components of acceleration in (\ref{a2}) and (\ref{a1}) yields 
\begin{eqnarray}
\ddot r+\frac{GM-\eta\kappa\cos^2\alpha}{r^2}-r\dot\phi^2=0,  \notag \\
\partial_t (r^2\dot\phi)+\frac{(1-\eta)\kappa\sin\alpha\cos\alpha}{r}=0.
\end{eqnarray}
Recalling that $\alpha=\sin^{-1}(v^{\phi}/c)$, we can write this up to
leading order in $v^{\phi}/c\ll 1$ as 
\begin{eqnarray}
\ddot r+\frac{GM-\eta\kappa}{r^2}-r\dot\phi^2=0,  \notag \\
\partial_t (r^2\dot\phi)+\frac{(1-\eta)\kappa v^{\phi}}{cr}=0.
\end{eqnarray}
Since $v^{\phi}=r\dot\phi$, we can integrate the second equation to give 
\begin{equation}
\dot\phi=\frac{v_0 r_0}{r^2}-\frac{\kappa (1-\eta)(\phi-\phi_0)}{cr^2},
\end{equation}
where $r_0$, $\phi_0$ and $v_0$ are the radial and angular positions and
speed of the solar sail at closest approach to the sun.

\section{Escape trajectories}

We will take the initial conditions to be at closest approach, where the
solar sail is deployed. Perihelion distances as small as 0.02 AU- 0.1 AU may be feasible for solar sails in the near future. For example, a trajectory design for a solar probe was presented in \cite{Guo} that includes repeated pole-to-pole sun flybys at a perihelion of slightly less than 0.02 AU.

Before the solar sail is deployed at the distance of closest approach $r=r_0$, the gravitational attraction of the sun causes the speed of the spaceship to increase as it gets closer to the sun. The Helios deep space probes would have traveled at the record speed of about 70 km/s
at 0.3 AU. This enables us to extrapolate that the following sampling of speeds $%
v_0$ are feasible for the near future: $v_0=133$ km/s at $r_0=0.1$ AU, $%
v_0=188$ km/s at $r_0=0.05$ AU and $v_0=298$ km/s at $r_0=0.02$ AU. 
We will use these as sets of initial conditions in order to demonstrate the Poynting-Robertson effect, though of course our orbital equations can be applied to any initial heliocentric distance and velocity. We take the conservative value of $\sigma=0.001$ kg/m$^2$ for the mass per area of the solar sail, since this could be large enough to take into account the mass of the load that is being transported. 

\begin{figure}[ht]
   \epsfxsize=3.3in \centerline{\epsffile{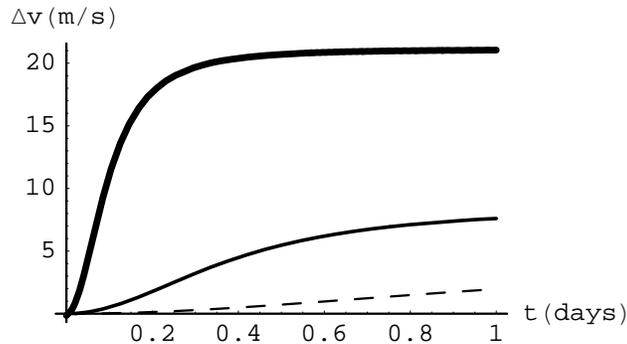}}
   \caption[FIG. \arabic{figure}.]{\footnotesize{The decrease in speed $\Delta v$ due to Poynting-Robertson effect for a solar sail initially moving at $0.02$ AU with a tangential speed of 100 km/s (dashed line), 300 km/s (solid line) and 600 km/s (bold line).}}
\label{fig3}
\end{figure}

Most of the acceleration for a solar sail takes place during the first day after it has been deployed. For our three sets of initial conditions, the decrease in speed $\Delta v$ due to the Poynting-Robertson effect for the first day of the voyage is shown in the Figure \ref{fig3}. 

\begin{figure}[ht]
   \epsfxsize=3.3in \centerline{\epsffile{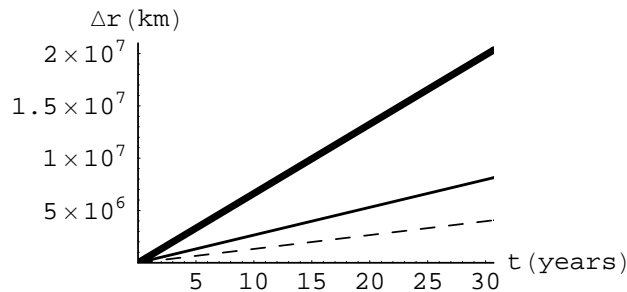}}
   \caption[FIG. \arabic{figure}.]{\footnotesize{The decrease in heliocentric distance $\Delta r$ due to Poynting-Robertson effect versus the duration of the voyage for the following sets of initial conditions: $v_0=133 $ km/s at $r_0=0.1$ AU (dashed line), $v_0=188$ km/s at $r_0=0.05$ AU (solid line), and $v_0=298$ km/s at $r_0=0.02$ AU (bold line).}}
\label{fig4}
\end{figure}

For a solar sail deployed at $0.02$ AU, the cruising speed after a 30-year voyage would be about 314 km/s. Throughout almost all of the voyage, the speed is about $\Delta v=$20 m/s less than what it would have been in the absence of this effect, which then has a cumulative effect on the heliocentric distance. The Poynting-Robertson effect decreases the heliocentric distance by an amount of $\Delta r\approx$ 20 million kilometers after a 30-year voyage. For a solar sail deployed at $0.05$ AU and $0.1$ AU, the heliocentric distance is lessened by about 7 million kilometers and 4 million kilometers, respectively, compared to what it would have been in the absence of the Poynting-Robertson effect. $\Delta r$ is shown as a function of the voyage time in Figure \ref{fig4} for our three sets of initial conditions.

\section{Bound orbits in the plane of the sun}

We will now consider the Poynting-Robertson effect on a solar sail-propelled satellite in a bound orbit. As an example, we take the effective areal mass to be $\sigma=0.00111$ kg/m$^2$. Table 1 lists the percentage decrease in the heliocentric distance after one year for a solar sail directly facing the sun and in a bound orbit at various initial distances from the sun. 

\begin{figure}[ht]
   \epsfxsize=2.5in \centerline{\epsffile{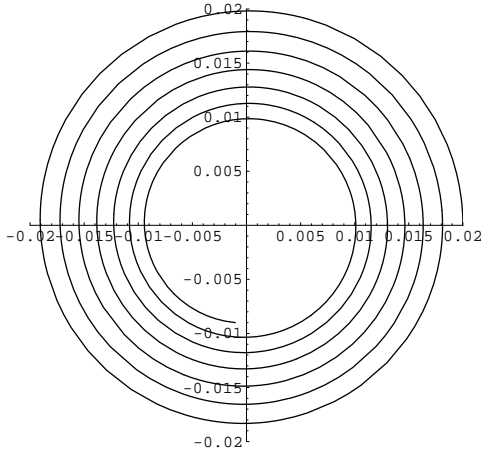}}
   \caption[FIG. \arabic{figure}.]{\footnotesize{Decrease in heliocentric distance $r$ due to the Poynting-Robertson effect for a solar sail with $\sigma=0.00111$ kg/m$^2$ that is initially in a circular orbit at $0.02$ AU.}}
\label{fig5}
\end{figure}

\begin{center}
\begin{tabular}{|c|c|c|c|}
\hline Initial Speed & Initial Distance & Distance After 1 Year & \% Decrease \\
\hline\hline 781.17 m/s & 0.38 AU & 0.37986 AU & 0.04\% \\
\hline 1076.76 m/s & 0.2 AU & 0.1985 AU & 0.8\% \\
\hline 1522.77 m/s & 0.1 AU & 0.096 AU & 4\% \\
\hline 2153.52 m/s & 0.05 AU & 0.043 AU & 14\% \\
\hline 2407.71 m/s & 0.04 AU & 0.027 AU & 33\% \\
\hline 2780.19 m/s & 0.03 AU & 0.0044 AU & 85\% \\
\hline\hline
\end{tabular}
\end{center}
\vspace{.6cm}
The initial tangential velocity is taken to be the value necessary for having a circular orbit in the absence of the Poynting-Robertson effect. As can be seen, the amount of change in the heliocentric distance increases dramatically as the initial heliocentric distance decreases. For a solar sail-propelled satellite initially around the orbit of Mercury at about 0.38 AU, the heliocentric distance decreases by only 0.04\% after one year. On the other hand, for an initial distance of 0.03 AU, the heliocentric distance decreases by about 85\% after one year. In the extreme scenario of a satellite beginning at $0.02$ AU, the heliocentric distance will decrease to less than $0.01$ AU in about 0.7 years, as shown in Figure \ref{fig5}. 

We will now consider a more general scenario in which the solar sail is tilted in the forward tangential direction (with respect to the velocity of the solar sail) at an angle $\psi_1$. Then the acceleration due to the portion of light that is reflected by the solar sail is given by
\be
a_{reflected}=\frac{(2\eta-1)\kappa}{r^2}\ \cos^2(\alpha+\psi_1)\,,
\ee
and is directed at an angle $\psi_1$ in the forward direction. The acceleration due to the portion of light that is absorbed is directed at an angle $\alpha$ in the backward direction and has a magnitude of
\be
a_{absorbed}=\frac{(1-\eta)\kappa\cos(\alpha+\psi_1)}{r^2}\,.
\ee
Projecting these two contributions to the acceleration along the tangential direction gives
\be
a^{\phi}=\frac{\big[ (2\eta-1)\cos (\alpha+\psi_1)\sin\psi_1-(1-\eta)\sin\alpha\big] \kappa\cos (\alpha+\psi_1)}{r^2}\,.
\ee

In principle, one could use the contribution to the acceleration in the forward tangential direction due to the reflected radiation to counteract the drag due to the absorbed radiation. In order to maintain a circular orbit at constant speed, we set $a^{\phi}=0$, which gives the condition
\be\label{psi-condition}
(2\eta-1)\cos(\alpha+\psi_1)\sin\psi_1=(1-\eta)\sin\alpha.
\ee
Since $\alpha=\sin^{-1}(v/c)$, this relates the tilting angle $\psi_1$ with the speed $v$ and reflectivity parameter $\eta$ of the solar sail. For $v\ll c$, we can approximate $\psi_1$ as
\be\label{psi-expansion}
\psi_1\approx \left( \frac{1-\eta}{2\eta-1}\right) \frac{v}{c}\,.
\ee
For example, in order to cancel the drag for the solar sail with an initial velocity of $3405.02$ m/s (and $\eta=0.85$), we find that $\psi_1\approx 2.4\times 10^{-6}$ rad. Since such precision for the tilting angle is clearly not realistic, one could instead perform periodic corrections on the tilting angle when the solar sail deviates too far from a circular orbit.

\section{Bound orbits out of the plane of the sun}

\begin{figure}[ht]
   \epsfxsize=2.5in \centerline{\epsffile{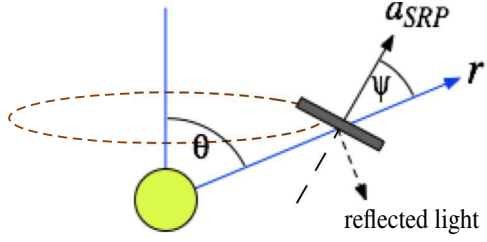}}
   \caption[FIG. \arabic{figure}.]{\footnotesize{In the absence of
the Poynting-Robertson effect, orienting a solar sail at an appropriate
pitch angle $\protect\psi$ in the polar direction would lead to a circular
orbit outside of the plane of the sun.}}
\label{fig6}
\end{figure}

We will now consider non-Keplerian orbits, for which the center-of-mass of the sun does not lie within the orbital plane \cite{mcinnes1,nonkepler1}. We orient our coordinate system such that a circular non-Keplerian orbit would correspond to constant $r$ and $\theta$, with the orbital radius given by $\rho=r \sin\theta$. In the absence of the Poynting-Robertson effect, a non-Keplerian orbit would be maintained with a suitable pitch angle $\psi_2$ in the $\theta$ direction and relative to the radial direction at the location of the solar sail, as shown in Figure \ref{fig6}. 

As before, due to the Poynting-Robertson effect, the solar radiation is at an angle $\alpha$ relative to the radial direction. We will consider a solar sail whose normal direction $\hat n$ to its surface is tilted at an angle $\psi_1$ in the $\phi$ direction, and which has a pitch angle $\psi_2$. We define $\Psi$ as the angle between $\hat n$ and the radial direction, $\Phi$ as the angle between $\hat n$ and the $\hat\phi$ direction, and $\gamma$ as the angle between $\hat n$ and the direction of the solar radiation. 
These angles are related to $\psi_1$, $\psi_2$ and $\alpha$ as follows:
\bea
\cos\Psi &=& \frac{1}{\sqrt{1+\tan^2\psi_1+\tan^2\psi_2}}\,,\nn\\
\cos\Phi &=& \frac{1}{\sqrt{1+\cot^2\psi_1+\tan^2\psi_2}}\,,\nn\\
\cos\gamma &=& \frac{1}{\sqrt{1+\tan^2(\alpha+\psi_1)+\tan^2\psi_2}}\,.
\eea

The fraction of light reflected $2\eta-1$ pushes the solar sail at an angle $\Psi$ relative to the radial direction, and the fraction of light absorbed $2(1-\eta)$ pushes the solar sail radially outwards. Thus, after performing the appropriate projections, we find
\bea\label{a3}
a^r &=& -\fft{G\td M}{r^2}\,,\nn\\
a^{\theta} &=& -\fft{(2\eta-1)\kappa \cos^2\gamma \sin\psi_2}{r^2}\,,\nn\\
a^{\phi} &=& \fft{(2\eta-1)\kappa\cos^2\gamma\cos\Phi-(1-\eta)\kappa\cos\gamma\sin\alpha}{r^2}\,,
\eea
where
\be
\td M\equiv M-\fft{\kappa}{G}\left[ (1-\eta)\cos\alpha+(2\eta -1)\cos\gamma\cos\Psi\right] \cos\gamma\,.
\ee
Acceleration in terms of spherical coordinates can generally be written as
\bea\label{a4}
a^r &=& \ddot r-r\dot\theta^2-r \sin^2\theta\ \dot\phi^2\,,\nn\\
a^{\theta} &=& r\ddot\theta+2\dot r\dot\theta-r\sin\theta\cos\theta\ \dot\phi^2\,,\nn\\
a^{\phi} &=& \fft{\sin\theta}{r} \partial_t (r^2\dot\phi)+2r\cos\theta\ \dot\theta\dot\phi\,,
\eea
where $\dot{}\equiv\partial_t$. Equating the components of acceleration in (\ref{a3}) and (\ref{a4}) yields the equations of motion for the solar sail:
\bea
\ddot r-r\dot\theta^2-r \sin^2\theta\ \dot\phi^2 &=& -\fft{G\td M}{r^2}\,,\nn\\
r\ddot\theta+2\dot r\dot\theta-r\sin\theta\cos\theta\ \dot\phi^2 &=& -\fft{(2\eta-1)\kappa \cos^2\gamma \sin\psi_2}{r^2}\,,\\
\fft{\sin\theta}{r} \partial_t (r^2\dot\phi)+2r\cos\theta\ \dot\theta\dot\phi &=& \fft{(2\eta-1)\kappa\cos^2\gamma\cos\Phi-(1-\eta)\kappa\cos\gamma\sin\alpha}{r^2}\,.\nn
\eea

\begin{figure}[ht]
   \epsfxsize=3.0in \centerline{\epsffile{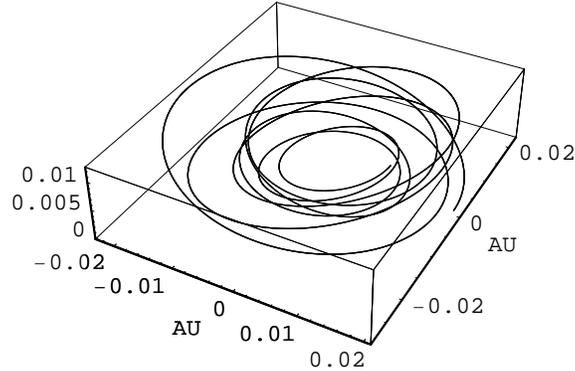}}
   \caption[FIG. \arabic{figure}.]{\footnotesize{A three-dimensional trajectory for a solar sail initially in a circular orbit outside of the plane of the sun at $0.02$ AU and a polar angle of $88.28^o$.}}
\label{fig7}
\end{figure}
\begin{figure}[ht]
   \epsfxsize= 5.5in \centerline{\epsffile{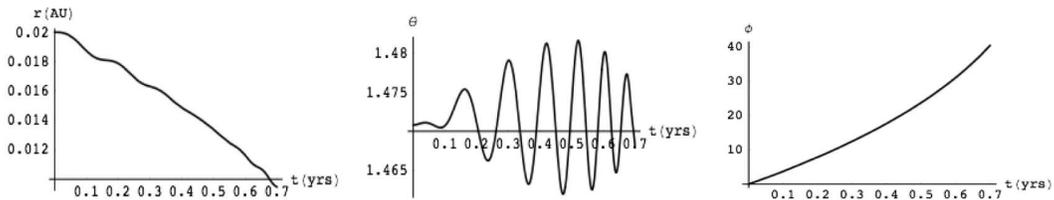}}
   \caption[FIG. \arabic{figure}.]{\footnotesize{Spherical coordinates of a three-dimensional trajectory for a solar sail initially in a circular orbit outside of the plane of the sun at $0.02$ AU and a polar angle of $84.27^o$.}}
\label{fig8}
\end{figure}

Figure \ref{fig7} shows a three-dimensional trajectory during one year, due to the Poynting-Robertson effect on a solar sail with $\sigma=0.00111$ kg/m$^2$ that is initially in a non-Keplerian orbit at $r_0=0.02$ AU and $\theta_0=88.28^o$ with an initial speed of $v_0=3406.63$ m/s and a pitch angle of $\psi=0.0054^o$. In order to depict the different types of motion that can occur in such a trajectory, Figure \ref{fig8} shows an example trajectory along the $r$, $\theta$ and $\phi$ directions for which the solar sail is initially at $r_0=0.02$ AU and $\theta_0=84.27^o$ with an initial speed of $v_0=3422.12$ m/s and a pitch angle of $\psi=0.0018^o$. 
As can be seen, the heliocentric distance $r$ is reduced to $0.01$ AU in $0.7$ years. Since the solar sail gets closer to the sun, the azimuthal velocity in the $\phi$ direction increases. Notice also that the solar sail exhibits oscillatory behavior in the $\theta$ direction, which can be minimized by tuning the pitch angle.

\section{Conclusions}

We have considered the Poynting-Robertson effect resulting from the
absorption of the solar radiation by a solar sail. The Poynting-Robertson
effect occurs at order $v^{\phi}/c$, and dominates over other special relativistic
effects which are at order $v^{2}/c^{2}$. For escape trajectories, this effect
decreases the cruising velocity as well as the heliocentric distance. For
the example of a solar sail with mass per area $\sigma=0.001$ kg/m$^2$ deployed between 0.02-0.1 AU, the Poynting-Robertson effect lessens the heliocentric distance after a 30-year voyage by 4-20 million kilometers.

As for bound orbits, if not compensated for, this effect can decrease the orbital speed of the solar sail, thereby causing it to slowly spiral towards the sun. We considered an example of a solar sail whose surface was directly facing the sun with mass per area $\sigma=0.00111$ kg/m$^2$. The percentage by which the heliocentric distance changes due to the Poynting-Robertson effect increases dramatically as the initial distance decreases. For an initial distance of 0.1 AU, the heliocentric distance decreases by about 4\% in one year, whereas an initial distance of 0.03 AU is decreased by 85\% in one year. In principle, the drag effect could be counter-balanced by an extremely small tilt of the solar sail in the polar direction. However, periodic adjustments when the solar sail deviates too much from the circular orbit are more feasible. 

We have also considered the Poynting-Robertson effect for a solar sail with a three-dimensional orbit. As an example, we have considered a solar sail with the necessary values of parameters so that, in the absence of the Poynting-Robertson effect, it would follow a closed circular orbit above the plane of the sun at a heliocentric distance of $0.02$ AU. While it spirals closer towards the sun due to the Poynting-Robertson effect, it also undergoes oscillatory motion in the polar direction, which can be controlled by the pitch angle.

For simplicity, we have worked in the approximation that the sun is a point-like light source. One could also take into account the angular variations of the solar radiation associated with the shape of the sun. However, this will not change our conclusion about the importance of the Poynting-Robertson effect.

Table 1: The percentage decrease in heliocentric distance due to the Poynting-Robertson effect for bound orbits with various initial conditions.

\end{document}